\documentclass[preprint,3p,twocolumn]{elsarticle}

\usepackage[utf8]{inputenc}
\usepackage{amsmath}
\usepackage{xspace}
\usepackage{graphicx}
\usepackage[bookmarksnumbered=true,bookmarksopen=true]{hyperref}

\newcommand{\dd}{\mathrm{d}}

\newcommand{\UNIT}[1]{\ensuremath{\,{\rm #1}}\xspace}

\newcommand{\MeV}{\UNIT{MeV}}
\newcommand{\GeV}{\UNIT{GeV}}

\newcommand{\fm}{\UNIT{fm}}
\newcommand{\fmc}{\UNIT{fm/c}}
\newcommand{\mb}{\UNIT{mb}}

\let\oldsqrt\sqrt
\def\sqrt{\mathpalette\DHLhksqrt}
\def\DHLhksqrt#1#2{%
\setbox0=\hbox{$#1\oldsqrt{#2\,}$}\dimen0=\ht0
\advance\dimen0-0.2\ht0
\setbox2=\hbox{\vrule height\ht0 depth -\dimen0}%
{\box0\lower0.4pt\box2}}

\journal{Physics Letters B}

\begin{document}

\begin{frontmatter}

\title{Investigating in-medium properties of the $\omega$ meson via the $\omega\rightarrow\pi^0\gamma$ decay}

\author[tp]{Janus Weil\fnref{fn1}}
\author[tp]{Ulrich Mosel}
\author[ep]{Volker Metag}

\address[tp]{Institut f\"ur Theoretische Physik, Universit\"at Giessen, D-35392 Giessen, Germany}
\address[ep]{II. Physikalisches Institut, Universit\"at Giessen, D-35392 Giessen, Germany}

\fntext[fn1]{Email: \href{mailto:janus.weil@theo.physik.uni-giessen.de}{janus.weil@theo.physik.uni-giessen.de}}


\begin{abstract}
  We investigate the feasibility of studying in-medium properties of the $\omega$ meson in photoproduction experiments via the decay $\omega\rightarrow\pi^0\gamma$. We use the GiBUU transport model to compare different methods of obtaining in-medium information, such as the invariant mass spectrum, transparency ratio, excitation function and momentum spectrum. We show that the final-state interaction of the pion poses a major obstacle for the interpretation of the invariant mass spectrum. The other three observables turn out to be fairly independent of final-state interactions and thus can give access to the $\omega$'s in-medium properties.
\end{abstract}


\end{frontmatter}


\section{Introduction}

The in-medium modification of vector-meson properties has been a topic of interest for quite some time now. Their vacuum properties are well-known nowadays, but the question of how their masses and widths change when embedded in a hadronic medium (either cold nuclear matter or the hot fireball of a heavy-ion collision) is debated heavily.
For recent reviews on in-medium effects, see \cite{Leupold:2009kz,Hayano:2008vn,Rapp:2009yu}.

The popular predictions on the one hand include standard many-body effects, like a collisional broadening of the meson spectral function due to collisions with the hadronic medium.
On top of that, a second class of predictions claims that the vector-meson masses
are shifted in the medium due to the (partial) restoration of chiral symmetry.
QCD sum rules can constrain these effects, but do not provide definitive predictions \cite{Leupold:1997dg}.

When studying in-medium effects of the $\omega$ meson by observing its decay products one has to remember that the invariant mass reconstructed from the four-vectors of the decay products always contains a product of spectral function and branching ratio. In choosing the decay channel
one has the choice between the rare dilepton decay mode, which is free of final-state interactions, and the more prominent hadronic or semi-hadronic decays, like $\omega\rightarrow\pi^0\gamma$. While the latter has the advantage of a much larger branching ratio of 8.3\% (roughly three orders of magnitude above the dilepton channel), it suffers from the fact that of one the decay products (namely the $\pi^0$) undergoes strong final-state interactions (FSI). For a long time it was commonly assumed that there are ways to cope with this issue, but we will show here that the pion FSI poses a major obstacle and in fact makes the $\pi^0\gamma$ decay unsuitable for certain in-medium studies.

Experimentally, the decay $\omega\rightarrow\pi^0\gamma$ has been studied intensively by the CBELSA/TAPS
collaboration in photoproduction reactions on nuclei \cite{Trnka:2005ey,Kotulla:2008aa,Nanova:2010sy,Nanova:2010tq}. Moreover, $\omega$ mesons in cold nuclear matter have been investigated via the dilepton decay channel by E325 at KEK \cite{Naruki:2005kd}, CLAS at JLAB \cite{Nasseripour:2007aa,Wood:2010ei} and HADES at GSI \cite{Agakishiev:2012vj}. While E325 claimed an omega mass shift without any broadening, the CLAS data indicate a broadening of the $\rho$ and $\omega$, but no shift. For the HADES data, no consistent interpretation is available yet regarding the $\omega$ meson: So far the p+Nb data only seem to show indications for an absorption of the $\omega$ meson. Overall, the in-medium mass shift of the $\omega$ meson is still an open issue from an experimental point of view.

We will start this paper with a short description of features of the GiBUU model, which are relevant for the present study of $\omega$ photoproduction reactions on nuclei. This will be followed by an overview over several methods for determining the in-medium properties of $\omega$ mesons in cold nuclear matter. As a new method we propose a study of the $\omega$ meson's momentum distribution.


\section{The GiBUU transport model}

\label{sec:gibuu}

Our tool for the numerical simulation of $\omega$ production and decay is the
GiBUU hadronic transport model, which provides a unified
framework for various types of elementary reactions on nuclei as well
as heavy-ion collisions \cite{Buss:2011mx,gibuu}. This model takes care
of the transport-theoretical description of the hadronic
degrees of freedom in nuclear reactions, including the propagation,
elastic and inelastic collisions and decays of particles.

In GiBUU the spectral one-particle phase-space distributions, $F(x,p)$, of all particles are obtained by solving the coupled Kadanoff-Baym equations
\cite{KadBaym} for each particle species in their gradient-expanded form \cite{Botermans:1990qi}

\begin{equation} \label{eq:os-transp.9}
\mathcal{D} F(x,p) - \text{tr} \left\{ {\Gamma f},{\Re S^{\text{ret}}(x,p)}\right\}_{\rm pb} = C(x,p)~,
\end{equation}

with

\begin{equation}
\mathcal{D} F = \left\{p_0 - H,F\right\}_{\rm pb}~.
\end{equation}

Here $\{\ldots\}_{\rm pb}$ denotes a Poisson bracket, and the so-called backflow term (second term on the lhs of (\ref{eq:os-transp.9})) is usually neglected. Further, $f(x,p)$ is the phase-space density related to $F$ by

\begin{equation}
F(x,p) = 2 \pi g f(x,p) \mathcal{A}(x,p) ~,
\end{equation}

where $\mathcal{A}(x,p)$ is the spectral function of the particle\footnote{$\mathcal{A}$ is normalized as $\int_0^\infty \mathcal{A}(x,p) \dd p_0 = 1$.} and $g$ is the spin-degeneracy factor.
The quantity $\Gamma$ in the backflow term is the width of the spectral function,
and $S^{\text{ret}}(x,p)$ denotes the retarded Green's function.

Our model explicitly includes off-shell transport of particles with density-dependent spectral functions. In particular, equ.~(\ref{eq:os-transp.9}) is compatible with the off-shell equations of motion obtained in \cite{Cassing:1999wx}, which leads to the correct asymptotic spectral functions of particles when they leave the nucleus. The expression $C(x,p)$ on the rhs of (\ref{eq:os-transp.9}) denotes the collision term that couples all particle species; it contains both a gain and a loss term. For a short derivation of this transport equation and further details we refer the reader to \cite{Buss:2011mx}. In order to solve the
BUU equation numerically, we rely on the test-particle ansatz. Here
the phase-space densities are approximated by a large number of
test particles, each represented by a $\delta$-distribution in
coordinate and momentum space.


The photoproduction of $\omega$ mesons is treated as described in Appendix C.8 of \cite{Buss:2011mx}, i.e. with parameterized matrix elements for $\gamma N\rightarrow \omega N$ and $\gamma N\rightarrow \omega\Delta$, which have been fitted to experimental data.

The absorption and rescattering of $\omega$ mesons in nuclear matter proceeds via the channels $\omega N\rightarrow \omega N$, $\omega N\rightarrow \pi N$, $\omega N\rightarrow2\pi N$ and $\omega N\rightarrow R$ (where R is a nucleon resonance). It is described in Appendix B.3 of \cite{Buss:2011mx}, just like $\pi N$ scattering, which reproduces the available data reasonably well. Following ref.~\cite{Muhlich:2006ps} we introduce an extra modification factor $K$ for the inelastic $\omega N$ scattering:

$$\tilde{\sigma}_{\omega N}^{inel}=\sigma_{\omega N}^{inel} \cdot K~.$$
For $K>1$, the additional contributions to the cross sections are put into the channel $\omega N\rightarrow2\pi N$.

\begin{figure}[t]
 \begin{center}
  \includegraphics[width=0.5\textwidth]{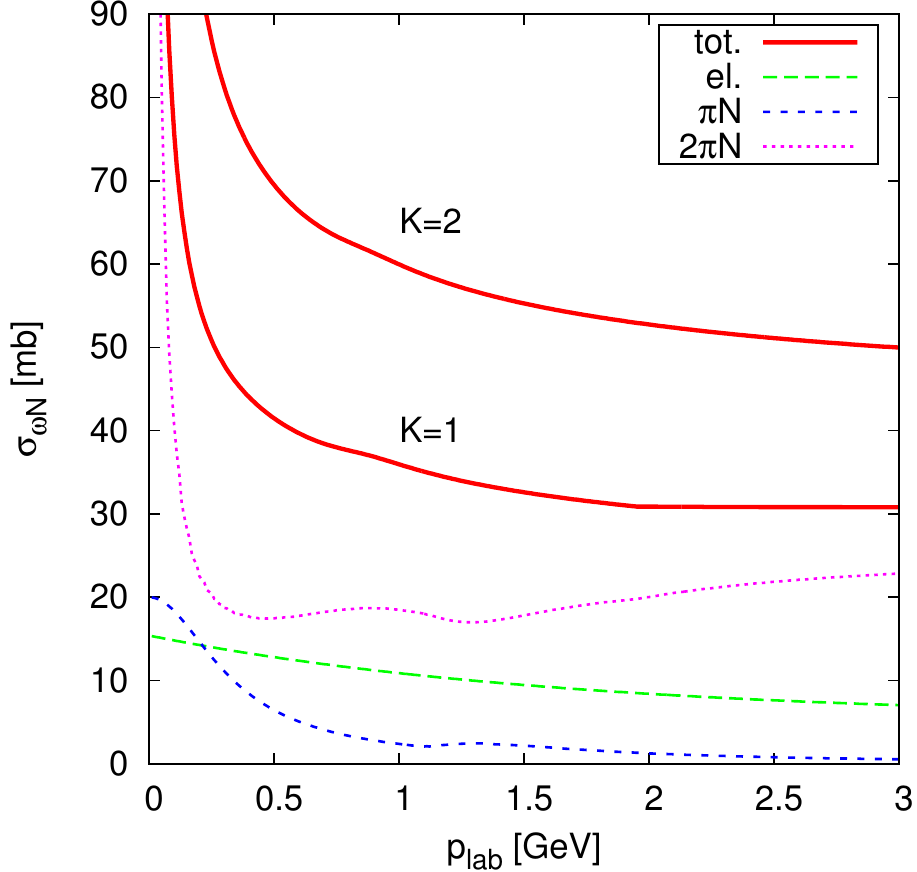}
 \end{center}
 \caption{(Color online) Total $\omega N$ collision cross section for K=1 and K=2.}
 \label{fig:xs}
\end{figure}

Fig.~\ref{fig:xs} shows the total $\omega N$ cross section for $K=1$ and $K=2$, as well as the contributions from different channels (for $K=1$ only).

We have chosen to put the additional absorption strength connected to the $K$ factor into the $2\pi N$ channel, since the $\pi N$ channel is constrained by the inverse reaction $\pi N\rightarrow\omega N$ via detailed balance. This strength could also be due to higher multiplicity channels like $\omega N\rightarrow3\pi N$, or could even mimic many-body absorption processes such as $\omega N N\rightarrow X$.

Our simulation procedure uses the full photon-energy range of 1.2 to 2.2\GeV with the appropriate $1/E_{\gamma}$ weighting factor of the Bremsstrahlung spectrum. Any in-medium changes, such as that of the mass and the width, are assumed to be linearly proportional to the nuclear density.

As far as in-medium modifications of the $\omega$ spectral function are concerned, we treat two different cases:

1. An in-medium mass shift according to

\begin{equation}
 m^*(\rho) = m_0 \left(1-\alpha\frac{\rho}{\rho_0}\right),
\end{equation}

with a scaling parameter parameter, $\alpha = 0.16$, as suggested in \cite{Hatsuda:1991ez}, corresponding to a 16\% mass shift at $\rho=\rho_0$. Here, $m_0=782\MeV$ is the vacuum mass of the $\omega$ meson and $\rho_0=0.168\fm^{-3}$ is the nuclear saturation density.

2. An increase of width due to collisions in the medium. The collision rate is given by 
\begin{equation}
 \frac{1}{\tau_{coll}}=\rho v_{rel}\sigma_{\omega N},
\end{equation}
where $v_{rel}$ is the relative velocity of the $\omega$ with respect to the hadronic medium and $\sigma_{\omega N}$ is given by the cross section depicted in Fig.\ \ref{fig:xs}. In order to maintain consistency between this collision rate and the collisional width appearing in the spectral function of the $\omega$ we use in the latter the momentum averaged expression
\begin{equation}
 \Gamma_{coll}(\rho)=\rho\left<v_{rel}\sigma_{\omega N}\right>=\Gamma_0\frac{\rho}{\rho_0}~.
\end{equation}
The value of $\Gamma_0=150\MeV$ is chosen to match the GiBUU collision term with $K=2$, and is compatible with the CBELSA-TAPS transparency-ratio data \cite{Kotulla:2008aa}.

We will show our results for these two cases separately and also for a combination of both. We assume a linear density dependence for both mass and width, since photoproduction experiments are anyway limited to small densities $\rho\leq\rho_0$, and pion FSI further reduces the sensitivity to large densities, as we will show in the following. Therefore, nonlinearities which can appear at higher densities can be neglected.


\section{Invariant Mass Spectrum}

\begin{figure*}[ht!]
 \begin{center}
  \includegraphics[width=\textwidth]{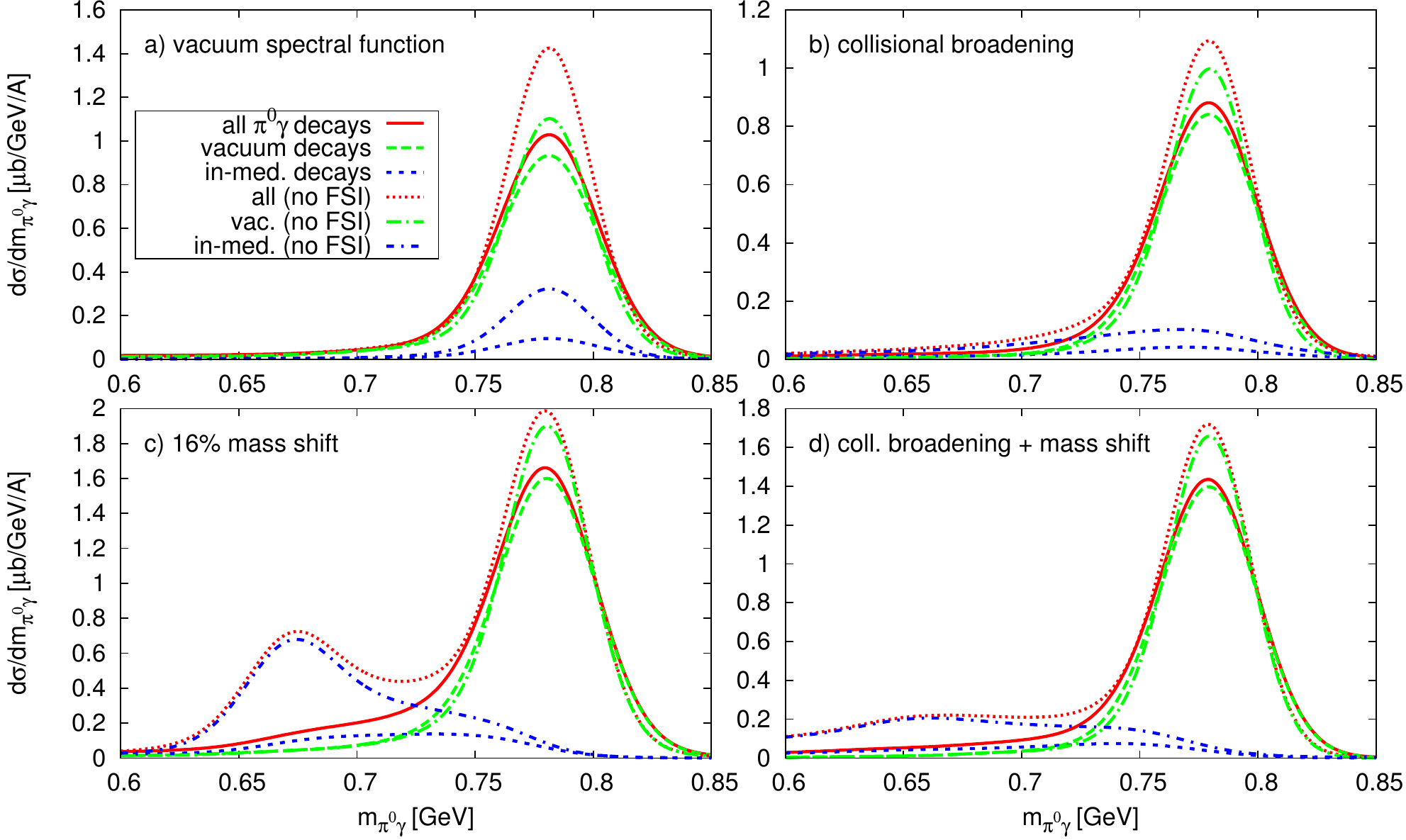}
 \end{center}
 \caption{(Color online) Calculated $\pi^0\gamma$ mass spectrum for $\gamma$+Nb at 0.9 - 1.3\GeV, in four different in-medium scenarios. Shown are the total spectrum as well as the contributions from in-medium ($\rho>0.1\rho_0$) and vacuum ($\rho<0.1\rho_0$) decays, with and without pion FSI.}
 \label{fig:mass1}
\end{figure*}

The idea to study in-medium properties of the $\omega$ meson via the $\pi^0\gamma$ invariant mass spectrum was first proposed in an early exploratory study by Messchendorp et al.\ \cite{Messchendorp:2001pa}, and further studied theoretically in a subsequent transport investigation \cite{Muhlich:2003tj}.
The first measurement followed soon \cite{Trnka:2005ey} and indeed claimed to observe a shifted in-medium peak in the $\pi^0\gamma$ invariant mass spectrum.
This claim was withdrawn later, after follow-up analyses \cite{Nanova:2010sy,Nanova:2010tq} with systematic studies of the background could not confirm the finding.

The results of our present transport study indicate that about 30 \%  of all $\omega\rightarrow\pi^0\gamma$ decays occur in the medium (i.e.\ at densities above a threshold of $0.1\rho_0$). However, there are several effects which affect the actual observability of any in-medium effect.
First of all, one expects a significant 'melting' of the in-medium peak due to collisional broadening. From the transparency ratio measurement of Kotulla et al. \cite{Kotulla:2008aa} a collisional width of $\Gamma_{coll}\approx130-150\MeV$ was extracted, which exceeds the $\omega$ decay width in vacuum by more than a factor 10. This means that the in-medium peak should be considerably smeared out, making it much harder to observe \cite{Lehr:2000ua}.

Moreover, not all of the in-medium decays happen at full nuclear density $\rho_0$ in the center of the nucleus. Every nucleus has a diffuse surface with a slowly dropping density distribution. The fraction of $\omega$ mesons decaying at the surface will further smear out the in-medium peak, if we expect a density-dependent mass drop. This holds for any decay mode of the $\omega$ meson.

For the $\pi^0\gamma$ channel, an additional limitation of the usefulness of the invariant mass spectrum as an indicator for in-medium effects is the strong final-state interaction (FSI) of the $\pi^0$ daughter particle. Most previous studies concentrated on the fact that the background created by rescattered pions can be suppressed by kinetic energy cuts \cite{Messchendorp:2001pa}. Little attention was paid to the fact that the pion FSI selectively favors decays at small densities. If a $\pi^0$ from an $\omega$ decay scatters with a nucleon, there are two possibilities: Either the collision is inelastic, leading to an absorption of the $\pi^0$ ($\pi^0N\rightarrow NX$), possibly due to charge exchange (e.g.\ $\pi^0p\rightarrow\pi^+n$); or the collision is elastic ($\pi^0N\rightarrow\pi^0N$), which typically causes an energy loss of the pion in the laboratory, so that the reconstructed $\pi^0\gamma$ invariant mass is changed dramatically. In both cases the event is basically unusable for
the $\omega$ reconstruction, since the original $\omega$ in-medium mass is lost.

The FSI has dramatic consequences for the in-medium part of the $\pi^0\gamma$ mass spectrum. Decay products coming from high densities at the center of the nucleus have little chance of making it outside without rescattering, so that mostly those from the surface will be observed, which carry little information on medium modifications. This means that the pion FSI cuts away most of the in-medium peak, as can be seen in Fig. \ref{fig:mass1}, and even in the very unlikely scenario of a pure mass shift, only a minor in-medium contribution to the spectrum is left. In the scenarios including collisional broadening, the in-medium peak is already broadened so strongly, that even without any FSI of the decay products, it would be hard to unambiguously measure a medium modification of the spectrum and to distinguish it from the background.

Although a much smaller effect of pion FSI was obtained in \cite{Das:2011gm}, this calculation also supports our conclusion that the $\pi^0\gamma$ invariant mass distribution is not very sensitive to in-medium modifications of the $\omega$ meson.

Note that Fig. \ref{fig:mass1} includes the mass resolution of the CBELSA/TAPS detector, incorporated by folding the spectrum with a Gaussian distribution of 19 MeV width. However, we have verified that the mass resolution of the detector is not a limiting factor, and that even an improvement of this resolution would not dramatically increase the chances of being able to observe any in-medium modification.

\begin{figure}[t]
 \begin{center}
  \includegraphics[width=0.5\textwidth]{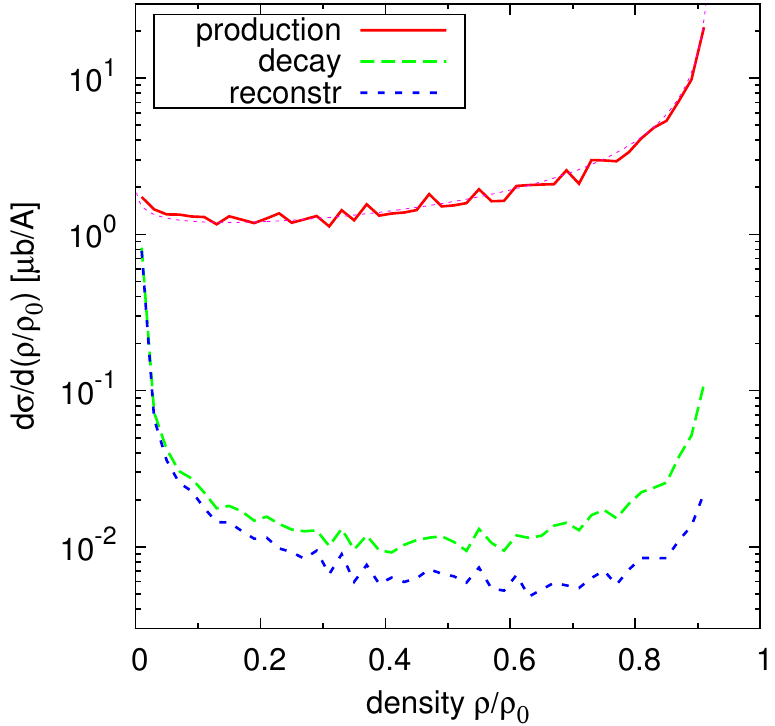}
 \end{center}
 \caption{(Color online) Distribution of the production points of $\omega$ mesons in photoproduction on Nb, $E_\gamma=0.9-1.3\GeV$, as a function of density (topmost line). Also shown are the $\omega\rightarrow\pi^0\gamma$ decay points without (middle line) and with (lowest line) pion FSI.}
 \label{fig:dens}
\end{figure}

The density distributions of the production and decay point (with and without FSI) are shown in fig.~\ref{fig:dens}. While the production density has a clear peak at full nuclear density, the decay point tends to lie in the vacuum due to the rather long lifetime of the $\omega$ meson ($\tau\approx23.2\fmc$). On top of this, the FSI of the decay pion further decreases the sensitivity to high densities. Additional cuts on the pion kinetic energy or the $\pi^0\gamma$ invariant mass can further reduce the reconstructed events from high densities.

In summary, we conclude that the $\pi^0\gamma$ decay channel is clearly inferior for invariant mass analyses compared to e.g. $e^+e^-$, mainly due to pion FSI. Moreover, we note that in general mass spectrum analyses are better suited for more short-lived mesons (like the $\rho$), which have a larger probability of decaying in the medium.


\section{Transparency Ratio}

Another way of investigating the in-medium properties of the $\omega$ meson is to measure its absorption as a function of nuclear mass number $A$. This method only gives information on the collisional width, but not on a possible mass shift. It does not require the measurement of actual in-medium decays, as is the case for the mass spectrum, but can rely completely on the four-momenta of the final decay products in vacuum. However, since pion FSI may decrease the number of finally reconstructed $\omega$ mesons the transparency itself does not give directly the attenuation of $\omega$ mesons alone, but has to rely on state-of-the-art transport calculations to extract the actual in-medium width. Only in the case of large collisional widths the pion FSI become negligible \cite{Kaskulov:2006zc}.

The so-called transparency ratio (normalized to ${}^{12}C$) is defined as:

\begin{align}
 T_A = \frac{12\,\sigma_{\gamma A\rightarrow\omega X}}{A\,\sigma_{\gamma {}^{12}C\rightarrow\omega X}}~.
\end{align}

The first measurement of this quantity was performed by the CBELSA/TAPS collaboration via $\pi^0\gamma$ decays in photoproduction experiments \cite{Kotulla:2008aa}. By assuming a low density approximation, an inelastic $\omega N$ cross section of about $70\mb$ and its momentum dependence were deduced from the measured transparency.

More recently, it was claimed in \cite{Rodrigues:2011zzb} that the CBELSA/TAPS data can be described by a much lower total $\omega N$ cross section on the order of $30\mb$. However, we note that the treatment presented there assumed a fixed photon energy of 1.5\GeV for comparison with the CBELSA/TAPS data, instead of the full bremsstrahlung energy spectrum of $E_{\gamma}=1.2-2.2\GeV$ used in the experiment. Using a fixed photon energy might be sufficient to describe the momentum-integrated transparency ratio, but it will certainly distort the momentum dependence.

\begin{figure}[ht]
 \begin{center}
  \includegraphics[width=0.5\textwidth]{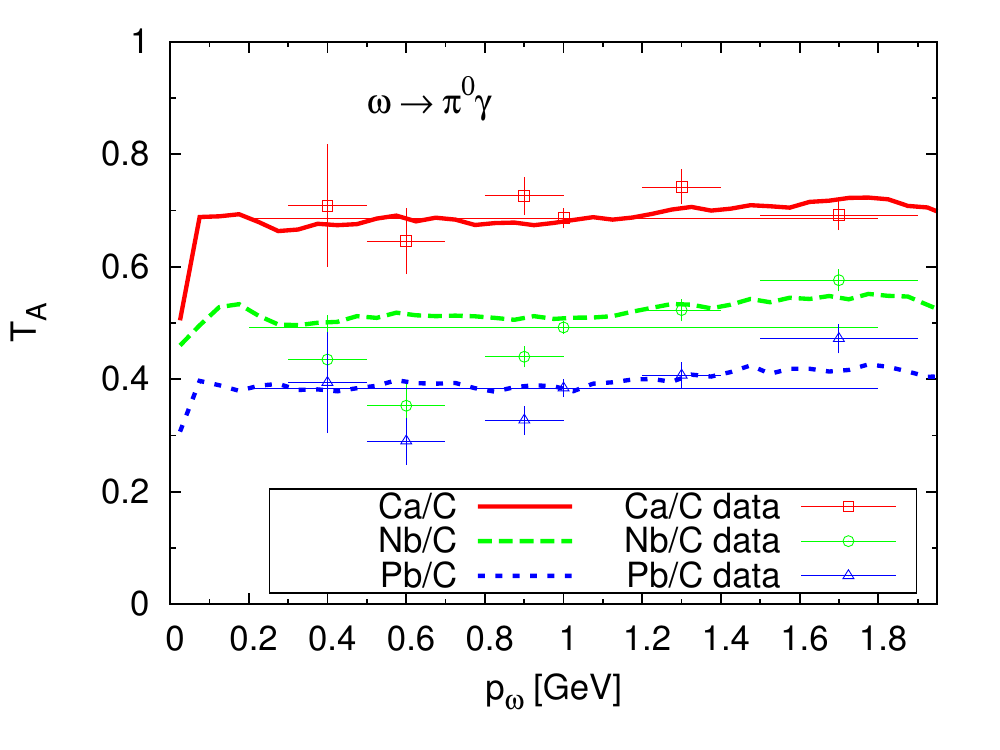}
  \includegraphics[width=0.5\textwidth]{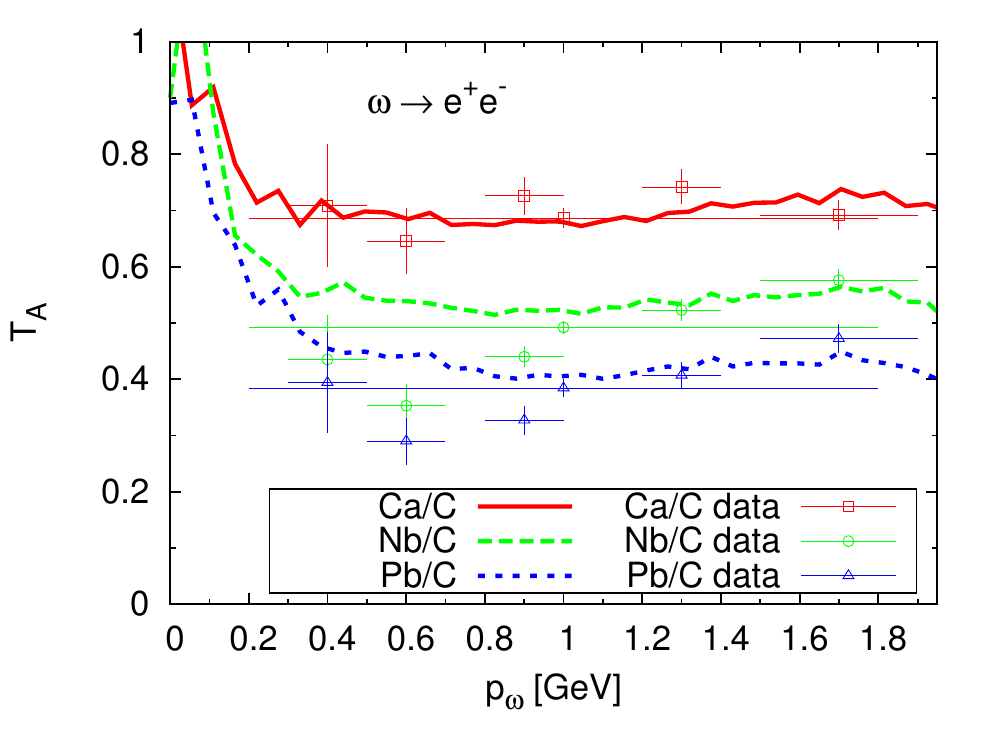}
 \end{center}
 \caption{Transparency ratio for three different nuclei relative to ${}^{12}C$, using $K=2$. Top: Obtained via the $\pi^0\gamma$ decay, with $T_{\pi}>150\MeV$, bottom: obtained from the dilepton decay, in comparison to the $\pi^0\gamma$ data from \cite{Kotulla:2008aa}.}
 \label{fig:tr}
\end{figure}

The results obtained with the GiBUU model are shown in Fig.~\ref{fig:tr}. To properly describe the data, we need to modify the inelastic $\omega N$ cross section by a factor of $K\approx2.0$. Our model yields a rather flat momentum dependence for the $\pi^0\gamma$ transparency ratio. Furthermore, it should be noted that the transparency ratio obtained via dilepton decays (as shown in fig.~\ref{fig:tr} bottom) is mostly compatible with the $\pi^0\gamma$ result, and only shows slight deviations at low momenta, which confirms that this observable is not strongly affected by pion FSI.

In fact, the $\omega$ transparency ratio has also been measured via dilepton decays from photon-induced reactions on nuclei by the CLAS collaboration at Jefferson Lab \cite{Wood:2010ei}. However, the CLAS data show a significantly larger absorption than the CBELSA/TAPS data, which currently can not be explained by any of the theoretical models (including ours).

Moreover, the dilepton data recently measured by the HADES collaboration \cite{Agakishiev:2012vj} could provide further constraints on the $\omega$ absorption in nuclear matter. An $\omega$ peak has been identified in both pp and pNb reactions at the same beam energy of 3.5 GeV. However, there are two obstacles for extracting the $\omega$ absorption cross section: First, the $\omega$ peak needs to be separated from an underlying background (mostly due to the $\rho$ meson), whose shape is nontrivial already in pp, and might be further modified in pNb \cite{Weil:2012ji}. Second, only a reference from pp collisions is available, and the $\omega$ production from pn collisions in not well constrained.


\section{Excitation Function}

It was argued in \cite{Muhlich:2006ps} that the $\omega$ excitation function, i.e. the total $\omega$ photoproduction cross section on a nucleus $d\sigma_{\gamma A\rightarrow \omega X}/dE_\gamma$, is sensitive to in-medium modifications of the $\omega$ meson. In particular, a downward mass shift of the $\omega$ in medium would lower the threshold for $\omega$ photoproduction and increase the cross section due to the enlarged phase space, which could be determined in an excitation function measurement.

\begin{figure}[hb]
 \begin{center}
  \includegraphics[width=0.5\textwidth]{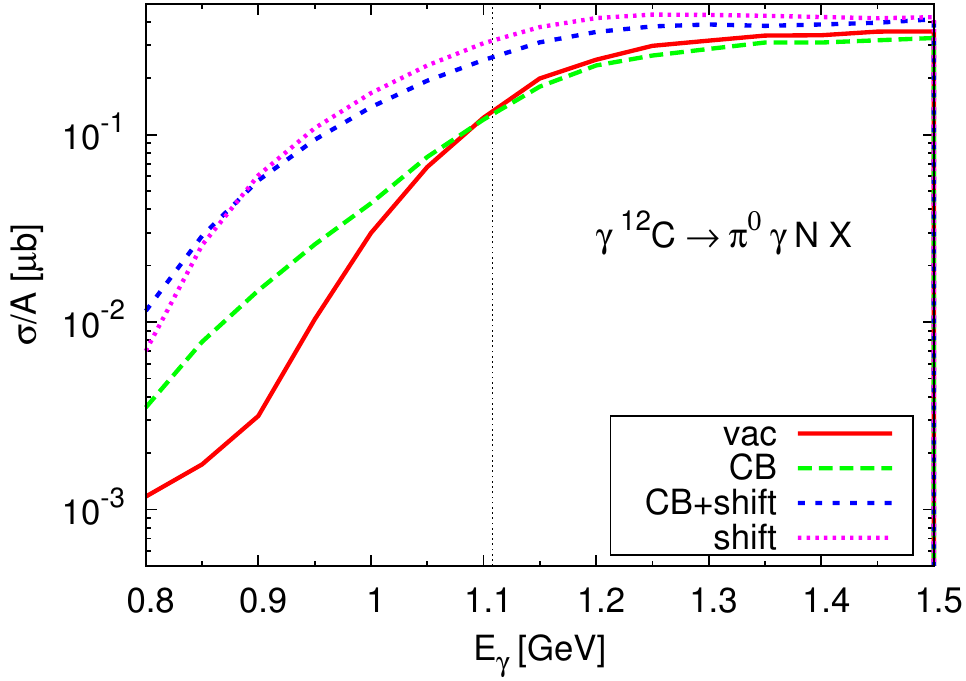}
  \includegraphics[width=0.5\textwidth]{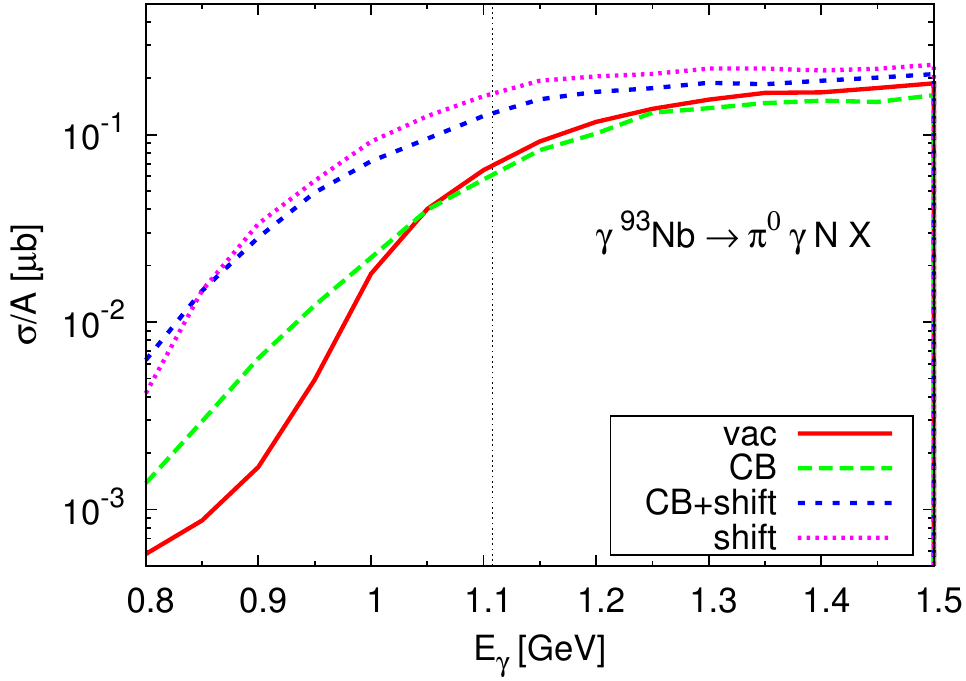}
 \end{center}
 \caption{(Color online) Calculated excitation function for photoproduction of $\omega$ mesons in the $\pi^0\gamma$ decay channel on ${}^{12}C$ and ${}^{93}Nb$ nuclei. The different curves represent the same in-medium scenarios as those used in fig.~\ref{fig:mass1}. The dashed line indicates $\omega$ photoproduction threshold on a free proton of $E_\gamma=1.108\GeV$.}
 \label{fig:exf}
\end{figure}

Fig.~\ref{fig:exf} shows the energy-dependent $\omega$ photoproduction cross section on C and Nb (observed through the $\pi^0\gamma$ decay) for photon energies from threshold up to 1.5\GeV. While the collisional broadening scenario only slightly enhances the subthreshold contributions (which presumably would be hard to measure, since the cross sections become very small there), the scenarios including a mass shift exhibit a much stronger enhancement, even above the free $\omega$ production threshold of $E_\gamma=1.108\GeV$.

It is important to note that the modification of the excitation function is proportional to the density at the production point, in contrast to the mass spectrum modifications discussed above, which are proportional to the density at the decay point.


\section{Momentum Spectrum}

In addition to the previously established methods discussed in the preceding sections, we propose here a new method for extracting in-medium properties of the $\omega$ meson from photoproduction data. Our simulations show that the momentum spectrum of $\omega$ mesons produced on a nucleus is sensitive to medium modifications of the $\omega$ properties.

This effect is due to the kinematics in the $\omega$ production process, which is altered by the $\omega$'s medium modifications. When an $\omega$ is produced with a lower mass, then, on average, it's total energy is lower. Furthermore, the meson has to come back to its free vacuum mass on its way out of the nucleus; the necessary energy is obtained by converting kinetic energy into mass. This conversion process lowers the momentum.

\begin{figure}[t]
 \begin{center}
  \includegraphics[width=0.5\textwidth]{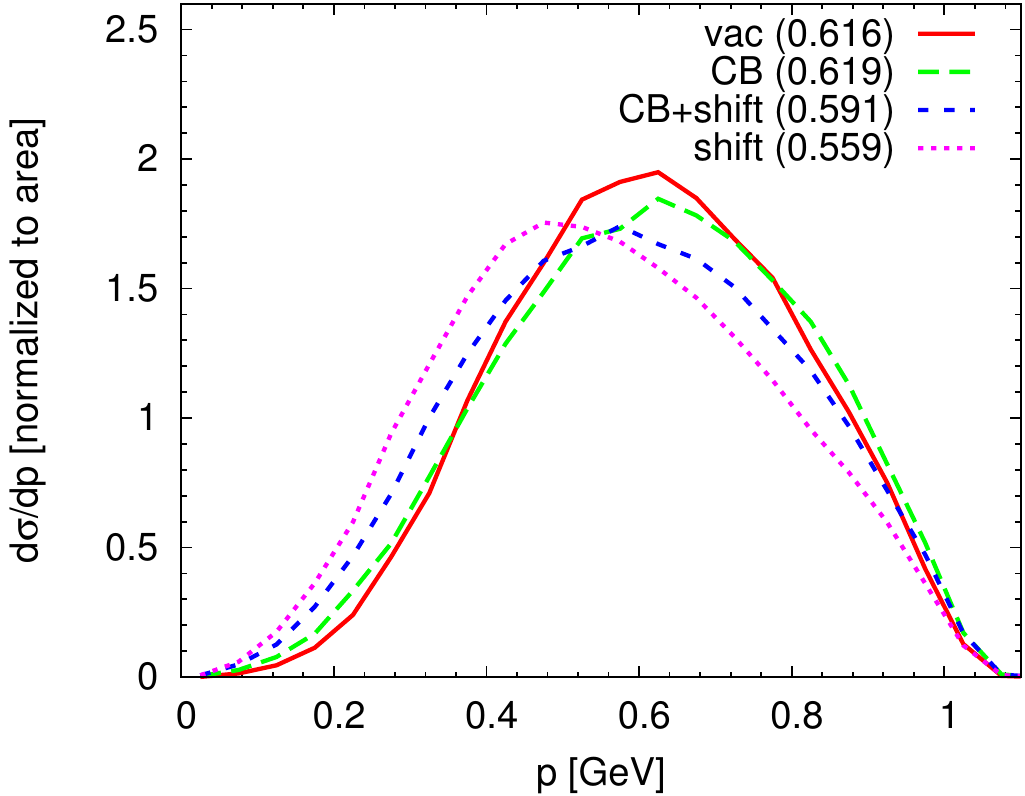}
 \end{center}
 \caption{(Color online) Calculated $\pi^0\gamma$ momentum distribution for $\gamma$+Nb at $E_\gamma=0.9 - 1.3\GeV$. The average momenta for the different in-medium scenarios are given in brackets (in GeV).}
 \label{fig:mom}
\end{figure}

Fig.~\ref{fig:mom} shows the momentum spectrum of a $\gamma$+Nb reaction with photon energies in the range of 0.9 - 1.3 \GeV. An invariant mass cut of $650\MeV<m<850\MeV$ was used to select the relevant $\omega$ mass region and to remove events with $\pi^0$ rescattering. It should be noted that the momentum distribution is quite sensitive to the angular distribution of the $\omega$ production process. The angular distributions used in the simulation properly reproduce the measurements on the free proton \cite{Barth:2003kv,Williams:2009aa}. For better comparison, the different curves are normalized to have the same integral. One can see that, while the collisional broadening scenario is hardly distinguishable from the vacuum curve, in the scenarios involving mass shift the momentum distribution is clearly shifted (the maximum of the distribution shows a clear offset and also the average is shifted by about 25 - 50 \MeV).

This means that a measurement of the $\pi^0\gamma$ momentum distribution might be used to determine the existence and magnitude of an in-medium mass shift of the $\omega$ meson. Preliminary data \cite{Metag:2011ji} seem to indicate that the in-medium scenarios involving mass shifts are disfavored.

\section{Conclusions}

We have evaluated various approaches for the determination of in-medium properties of the $\omega$ meson in cold nuclear matter via the $\pi^0\gamma$ decay in photoproduction experiments. Our analysis shows that the lineshape analysis of the $\pi^0\gamma$ invariant mass spectrum suffers mainly from the hadronic FSI of the pion, which drastically reduces the sensitivity to medium modifications. This means that even if the $\omega$ mass was shifted in the medium, it would be very hard to detect it via the $\pi^0\gamma$ mass spectrum.

Potentially larger effects could be expected from two other methods of determining the in-medium mass of the $\omega$: the excitation function and the momentum spectrum. Both are sensitive to the density at the production place of the meson, and not at the decay location (as is the case for the lineshape).

The transparency ratio, on the other hand, yields a reliable determination of the $\omega$ absorption cross section. It is not as sensitive to pion FSI, since it relies mostly on decays of $\omega$ mesons outside the medium. However, it can not make any statement about an in-medium mass shift of the $\omega$.


\section*{Acknowledgments}

Work supported by HIC-for-FAIR, HGS-HIRe and DFG under TR16. We thank Stefan Friedrich for the help in producing fig.~\ref{fig:dens}.


\end{document}